\pdfoutput=1 
\documentclass[twocolumn,showpacs]{revtex4-1}
\usepackage{graphicx}
\usepackage{dcolumn}
\usepackage{bm}
\usepackage{amssymb}
\usepackage{amsthm}
\usepackage{color}

\usepackage[           
a4paper,                
colorlinks=true,        
citecolor=green,        
linkcolor=cyan,         
menucolor=blue,         
urlcolor=magenta,       
bookmarks=false
]{hyperref}

\begin{document}

\title[Moral foundations and interacting neural networks]{Moral foundations in an interacting neural networks society}

\author{R. Vicente}
\email[E-mail me at: ]{rvicente@ime.usp.br}
\affiliation{Dept. of Applied Mathematics, Instituto de Matem\'atica e Estat{\'\i}stica, Universidade de S\~ao Paulo, 05508-090, S\~ao Paulo-SP, Brazil.}

\author{A. Susemihl}
\affiliation{Artificial Intelligence Group, Technical University Berlin, 
 Franklinstra{\ss}e, 28/29, D-10587 Berlin, Germany.}

\author{J.P. Jeric\'o}
\author{N. Caticha}
\email[E-mail me at: ]{nestor@if.usp.br}
\affiliation{Dep. de F{\'\i}sica Geral, Instituto de F{\'\i}sica, \\
Universidade de S\~ao Paulo, CP 66318, 05315-970, S\~ao Paulo-SP, Brazil.}

\begin{abstract}
The moral foundations theory supports that people, across cultures, tend  to consider a small number of dimensions when
classifying issues on a moral basis. The data also show that the statistics of weights attributed 
to each moral dimension is related to self-declared political affiliation, which in turn  has been connected to cognitive learning styles by
 recent literature in neuroscience and psychology.   Inspired by these data,  we propose a simple statistical mechanics model with
interacting neural networks classifying vectors and  learning from  members of their social neighborhood
 about their average opinion on a large set of issues. The purpose of learning is to reduce dissension among agents even when disagreeing.
We consider a family of learning algorithms parametrized by $\delta$, that represents the  importance given to corroborating (same sign) opinions. We define 
an order parameter   that quantifies the diversity of opinions in a group with homogeneous learning style. Using Monte Carlo simulations and 
a mean field approximation we find the relation between the order parameter and the learning parameter  $\delta$ at a temperature we 
associate with the importance of social influence in a given group. In concordance with data, groups that rely more strongly on corroborating 
evidence sustains less opinion diversity. We discuss predictions of the model and propose possible experimental tests. 
\end{abstract}

\maketitle

\section{\label{sec:intro} Introduction}

Sociophysics, the approach to mathematical modelling of social science,  is still maturing as a scientific field  \cite{GalamBook}. Opinion dynamics, voting, social influence and contagion models have been thoroughly studied \cite{Galam,Castellano}, patterns in social data have been identified (e.g. \cite{Bouchaud} or \cite{Fortunato} and references therein)  and  some successful predictions have been achieved (e.g. \cite{GalamQQ}).

In this paper our aim is to present a data driven  statistical mechanics model  for  the formation of opinions about morality. We would like to verify if we can explain  features of social data by considering  a stylized model for  neurocognitive processes  well described in the literature. Clearly practical limits to  such a goal have to be considered.  At the scale of individuals,  neurocognitive data inspiring any modeling are always exposed  to  ecological validity issues with multiple uncontrolled causes. At the social scale, we also have to keep in mind the sheer complexity of human nature and human
 relationships.  By stylized model we here mean a model to be used mainly  to connect  pieces of empirical evidence, to help the identification of important variables and as an aid to formulate new empirical questions. Furthermore, we would also like to have a model capable of making predictions after fitting a few key parameters  to empirical data.

We argue here and in our previous work \cite{Cavi10} that the evidence available on the moral classification problem can be accommodated by assuming agents that are conformist classifiers adapting to their social neighborhood by reinforcement learning. Empirical evidence regarding different cognitive styles can then be represented in the model as distinct learning algorithms following the now established tradition 
of the statistical mechanics of learning \cite{Engel}. Studies on social psychology \cite{Spears} allow the further simplification of assuming  that social influence only takes place between individuals perceived as similar. As a first approximation we thus assume that the social network can be partitioned into homogeneous social influence subnetworks, each one with a given cognitive style or learning algorithm.

But what do these conformist agents classify? We assume that any issue under debate can be parsed into a discrete set of independent attributes or dimensions.
 The modern theory of moral foundations \cite{Haidt1} suggests that, as far as morality is concerned, these dimensions are not many more than five, namely: (a) harm/violence;
 (b) justice/fairness; (c) in-group loyalty; (d) respect for authority; and (e) purity or sanctity. For our modeling effort it is, however, sufficient that morality can 
be parsed into a discrete number of identifiable dimensions. As a starting point we do not consider 
the origin of these dimensions, its particular
 meanings or the practical issues that may be involved in  trying to  parse a given subject into these dimensions.These five dimensions have been found empirically to be sufficient
to characterize political orientations along the liberal conservative spectrum.
The need for a sixth dimension, (f) liberty/oppression, has been included 
to extend the description to extend the spectrum to also include libertarians, but this is outside the scope of this article. 
 
We thus consider that the moral content of an issue may be represented by a direction in a unit radius $D$-dimensional hypersphere $\mathbf{x}\in \mathbb{S}^D$ . In the course of  daily social relationships an individual $j$ will
 be exposed to a variety  of issues of diverse
 moral content parsed  as  $\mathbf{x}_j^\mu$ with $\mu=1,2,\cdots$. For each of these issues an opinion $\sigma_j^\mu \in [-1,1]$ with a sign 
and an amplitude $|\sigma_j^\mu|$ is displayed. The sign can be interpreted as providing a for/against information and the amplitude
 as carrying information on how convict individual $j$ is. A way to describe
 a classification task of this sort is by assuming that $\sigma_j^\mu=\mathbf{x}_j^\mu\cdot\mathbb{J}_j$, where $\mathbb{J}_j$ is an adaptive internal 
representation, inaccessible  to other individuals, used by individual $j$ to perform moral classification tasks. For simplicity we will study the case where all  moral
 vectors are normalized to unit length $\mathbb{J}_j \in \mathbb{S}^D$. This also implies that differences in moral values are not interpreted as 
any type of moral superiority and that no  moral shallowness is implied by the differences. Thus only the direction  the moral vector points is considered as important, removing a layer of complexity in the interpretation of the model.

A conformist individual will then seek  agreement
 with social neighbors in moral classifications by adjusting internal representation $\mathbb{J}_j$. 
Employing the statistical mechanics of learning jargon, we are  supposing that model  agents are interacting {\it normalized linear perceptrons}  \cite{Engel} (for previous studies of interacting neural networks see \cite{Metzler00interactingneuralPRE,Vicente} ). 

The moral parsing of issues $\mathbf{x}_j^\mu$ is {\it subjective}, to say, two individuals would not necessarily agree on
  how a given issue should be parsed. To further simplify our model we suppose that conformist classifiers
 do not adapt to opinions about every issue separately, but, instead to a normalized average over a large set of $P$ issues: 
\begin{equation}
 h_j= \left(\frac{\sum_{\mu=1}^P\mathbf{x}_j^\mu}{\|\sum_{\mu=1}^P\mathbf{x}_j^\mu\|}\right) \cdot\mathbb{J}_j
\label{opinion_field}
\end{equation}
Assuming that there are no relevant biases or correlations in the individual parsing through the social network, we write the opinion field as  $h_j=\mathbb{Z}\cdot\mathbb{J}_j$, where the mean issue
\begin{equation}
\mathbb{Z}=\frac{\sum_{\mu=1}^P\mathbf{x}_j^\mu}{\|\sum_{\mu=1}^P\mathbf{x}_j^\mu\|},
\end{equation}
 is supposed to be {\it objective} (or independent of the index $j$). We therefore suppose that conformist agents classify the average issue represented by the vector $\mathbb{Z}$ and exchange information 
about their classifications in the form of opinion fields $h_j=\cos\theta_j$, where $\theta_j$ represents the angle between the internal (moral) representation 
$\mathbb{J}_j$ and a symmetry breaking direction $\mathbb{Z}$ given by the mean issue parsed into moral dimensions.

For the sake of brevity  we here only provide a short summary of empirical evidence and  focus on the statistical mechanics model. To the reader interested in knowing more about relevant empirical sources we suggest reading our previous work on the subject \cite{Cavi10}. Empirical evidence suggests that individuals are conformist agents that adapt to each other by reinforcement learning \cite{Klucharev}, that
cognitive styles are diverse \cite{Amodio} and that 
agents are more strongly influenced by other agents with similar style. Learning styles can be parametrized
 by $\delta \in [0,1]$ that represents the difference in how the agent
treats corroborating (agreement) information against how she deals with novelty (disagreement). Agents with larger $\delta$, weight disagreement and agreement more similarly, while those with smaller $\delta$ give more weight to 
novelty than to corroboration. 
Additionally, psychological and neurocognitive data suggest a positive correlation between cognitive style and 
self-declared political affiliation ({\it p.a.})\cite{Amodio}.

The aggregate behavior, represented by statistics of the opinion fields $h$, can be derived using statistical mechanics and then compared to 
social data on moral foundations. Given $\delta$, the model {\it predicts} the shape of histograms $p(h|\delta)$. If we postulate that cognitive style, e.g. 
$\delta$ positively correlates with political affiliation, 
the model also predicts certain aspects of the behavior of 
$ p(h|\mbox{\it p.a.})$. Alternatively, similarity between  $p(h|\delta)$ 
and $ p(h|\mbox{\it p.a.})$ leads to a confirmation that cognitive style is 
positively correlated to political affiliation.

The simultaneous comparison
of six predicted histograms $ p(h|\mbox{\it p.a.})$ ($\mbox{\it p.a.}=1,2,\cdots,6$) with data requires the selection of two phenomenological parameters: 
the average node degree in the social influence subnetwork $\bar{k}$ and the average social pressure per social neighbor $\alpha$. A mean field approximation predicts that histograms depend on the total social pressure, namely, on these two parameters combined as  $\bar{k} \alpha$.   An optimization procedure, designed to maximize the similarity between  predicted and empirical histograms, can then be used to estimate $\bar{k} \alpha$.

This paper is organized as follows. In the next section we describe the available data. Section \ref{sec:model}
 describes the statistical mechanics model in detail and an analysis  via Monte Carlo simulations. Section \ref{sec:mft} treats analytically a mean field version of the model. In Section\ref{sec:facebook} we simulate the model in a real-world social graph provided by the Facebook. The dynamical behavior of the model is discussed in section \ref{sec:dyn}. We discuss the meaning of diverse cognitive styles in the light of the model in section \ref{sec:discussion}.  Finally, a section with conclusions and perspectives is provided.

\section{\label{sec:empirical} Data on Moral Foundations}

In a series of papers \cite{Haidt1,Haidt04, Haidt4,Haidt2,Haidt07,Haidt3} Jonathan Haidt and coworkers 
have described moral foundation theory (MFT),
an heuristically driven theory dealing with the
foundations of moral psychology. Its aim is to understand
statistically significant differences in moral valuations 
of social issues and their
association to  coordinates of a political spectrum.

Following  Kohlberg \cite{Kohlberg} and  Gilligan \cite{Gilligan}
work in moral psychology in the western world tradition, dealt with 
the representation of moral issues in a two dimensional space.
The first historically identified dimension is
related to whether an action leads to harm and violence
or not. Later, the existence of a second dimension, associated to
justice and fairness was introduced.
By analysis of literature  extending across time, geography and  scientific disciplines, 
Haidt and coworkers introduced the  main ingredient that yields a
foundation theory \cite{Haidt1}: humans when
classifying issues as either moral or immoral, 
navigate not in a two dimensional space, but in one that is at least
five  dimensional. 
Valuations along these dimensions,  called foundations 
in the literature of moral psychology, are 
necessary to characterize the moral content of a given issue. 
 
Their striking quantitative result is extracted from massive amounts of
data: as the political spectrum is traversed from liberal to conservative, there is an increase, from 
two to five, in the number of  moral dimensions considered relevant 
to form opinions.
Liberals regard (a) harm/violence
 and (b)justice/fairness, the
previously identified dimensions, as the most relevant foundations. 
In addition, conservatives hold  (c) in-group loyalty, (d) respect for authority
and  considerations about (e) purity or sanctity in a 
considerable higher position than liberals.
That means that, independently of the semantic role of the attributes, 
it can be asserted that 
liberals rely on a different subset of moral foundations than
conservatives. We here employ liberal in the manner defined in the USA
as social liberal.   

The data we have analyzed were furnished by Jonathan Haidt \cite{Haidt1,Graham} \footnote{Raw data can be requested to Haidt's team at \href{http://www.yourmorals.org/}{http://www.yourmorals.org/}. Preprocessed data and source codes can be downloaded at
 \href{https://github.com/renatovicente/statmech_mft.git }{https://github.com/renatovicente/statmech\_mft.git.}} They were
collected from the answers to a specially designed 
 questionnaire aimed at probing opinions about 
morally relevant situations. 
For each of $N=14250$ respondents, Haidt and coworkers extracted 
five dimensional moral vectors with components 
related to five Moral Foundations. Each vector was labeled 
by the subject's self-declared political affiliation ({\it p.a.} ranges from
$=1$ (very liberal) to $7$(very conservative)).

\begin{widetext}
\begin{figure*}[!ht]		
\centering \includegraphics[width=0.6\textwidth]{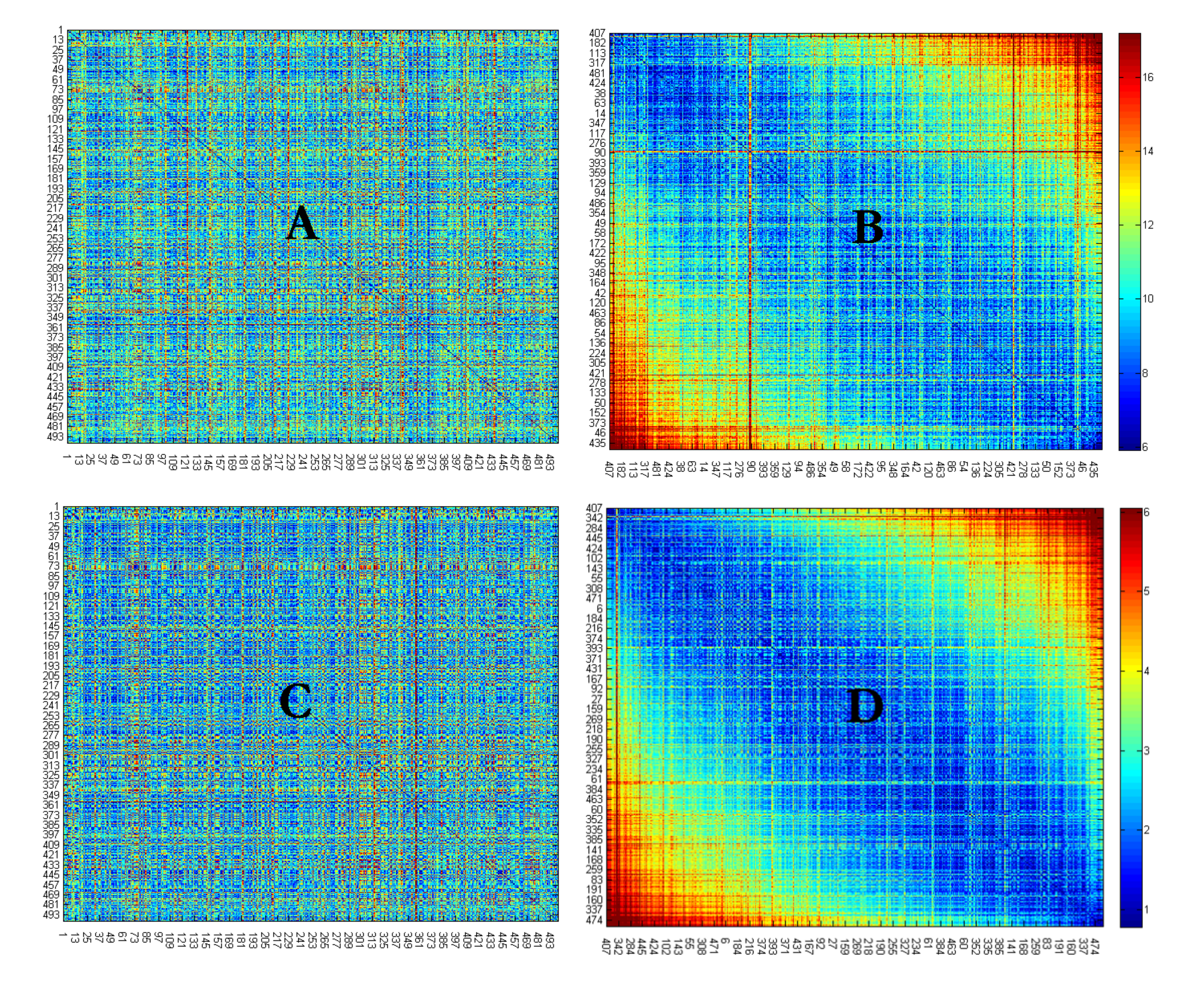}
\centering \caption{{\bf Experimental data - matrices of distances}: top row,(A) 30 dimensional representation of
 subjects, bottom row, (C) 5 dimensional space of moral foundations. 
Left column: before, Right column: (B) and (D), respectively, after 
application of SPIN algorithm. Blue means near and red far.}
\label{fig:figspin}
\end{figure*}
\end{widetext}

The questionnaires consisted of 30 questions probing the subject in the
5 moral dimensions. From the set of answers
the five dimensional vectors with components in the interval $[0,5]$ are extracted.
Thus a subject can either be represented as a point in the 30 dimensional
space of questions, or in the reduced moral foundation space. 
It is interesting to see if the cloud
of data points have a similar
structure in both spaces. A negative answer would be 
indicative that the reduction has either deleted or invented some structure. We stress that we are not looking for clusters of different
political affiliation in this analysis. The 
data should, if the questionnaires are relevant, characterize 
the relation between the complex moral valuation systems and 
the simple one dimensional continuous political affiliations.

The consistence of the five factors model has been already probed in \cite{Graham}.
We here confirm that the data reduction is significant 
employing a visualization technique know as SPIN 
\cite{Domany} used in the analysis of large dimensional
data sets in bioinformatics. It is a dimensional reduction 
technique that identifies a nonlinear one dimensional manifold 
irrespective of the embedding dimension of space.
  In both spaces we use an Euclidean
distance to measure how different are any two individuals. A permutation
of a set of individuals is done in order to give close labels to 
pairs that are close in the original space and try to give them 
far apart labels in case their distance is large. 
It has the advantage
that the shape of one dimensional  structures can be identified. Figure \ref{fig:figspin} shows 
distance matrices for balanced sets of subjects. Random samples of 
questionnaires were chosen keeping the same sample size for each category 
of political affiliation.
In the left column we show the matrices before the permutation, in the 
right column, after the permutation. In the top row, we show the bare
data from the 30 dimensional space. In the lower row, the data of
the reduced 5 dimensional space. 
The fact that the  one dimensional structure that
can be seen embedded in both spaces is similar, gives further support to the 
hypothesis that the 5 dimensional reduction to the moral foundations
matrix from the 30 dimensional questionnaires preserve a one dimensional
geometrical structure in the cloud of data points which is associated
to political affiliation.

\begin{widetext}
\begin{figure*}[!htb]		
\centering\includegraphics[width=0.9\textwidth]{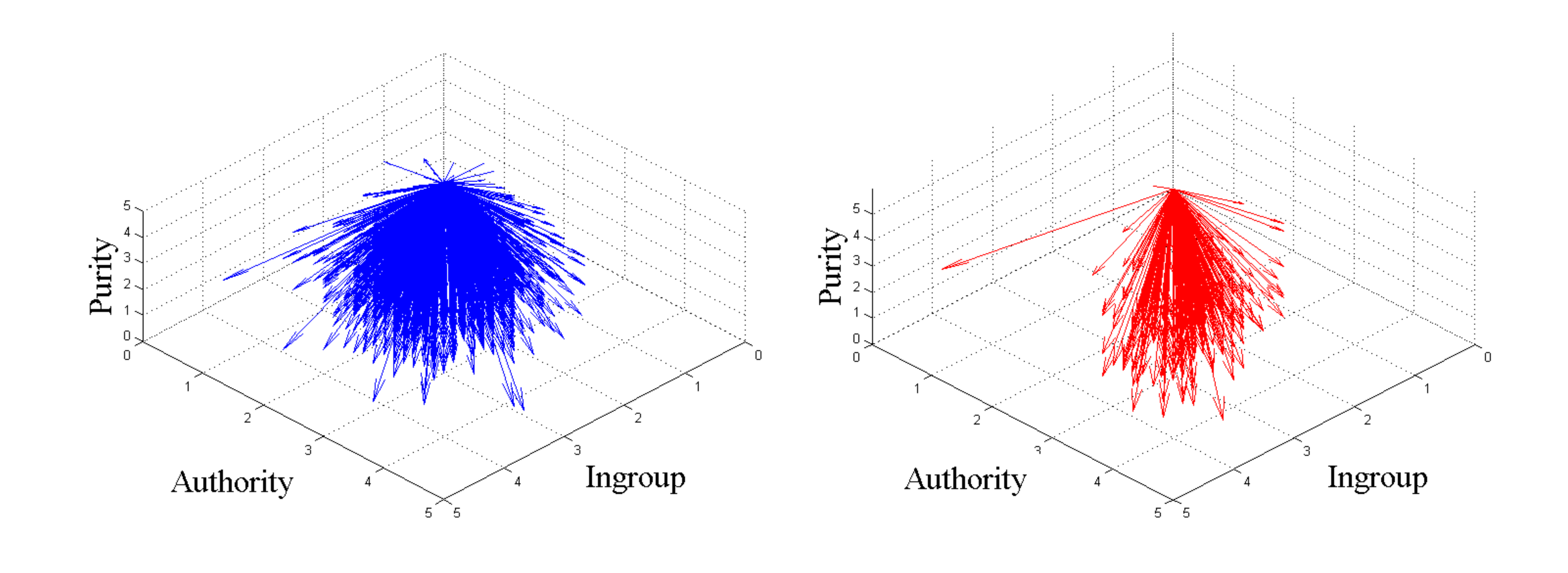}
\centering \caption
 {{\bf Empirical moral vectors}: Three-dimensional projections of moral vectors for very liberal ({\it p.a.}$=1$, blue) 
and conservative ({\it p.a.}$=6$, red) subjects. Axes are labeled according to the associated moral foundation.}
\label{fig:figmoral}
\end{figure*}
\end{widetext}

Figure \ref{fig:figmoral} depicts three components for {\it p.a.}=1 (very liberal) and {\it p.a.}=6 (conservative).  
For comparison purposes  
 moral vectors   $\mathbb{J}_i$ of the subjects were normalized to unit length as in the 
statistical mechanics model. In the model the vector $\mathbb{Z}$ is a symmetry breaking direction determined 
by the set of issues under discussion in a society.
This set is a complicated thing to define. In particular, we have no access to the
parsing that would permit its representation in five dimensions. To make possible a verification of the model,
 we have to identify the analogous of the direction $\mathbb{Z}$ within the data. Looking at Figure \ref{fig:figmoral} a natural choice, further justified by considering the model dynamics, 
consists on identifying $\mathbb{Z}$ to the average vector 
within the conservative ({\it p.a.}$=6$) and very conservative ({\it p.a.}$=7$) classes. 

We then calculate  empirical histograms $H_E$ for   ${h_j}=\mathbb{J}_j\cdot\mathbb{Z}$, that can be interpreted as the  effective number of moral dimensions of a  subject labeled by $j$.
These histograms $H_E$ characterize the different political groupings 
in a semantic free manner and will be compared to similar statistics obtained using analytical methods and numerical simulations.

\section{Statistical mechanics model}
\label{sec:model} 
Agents exchange information in the form of fields $h_j=\mathbb{J}_j \cdot \mathbb{Z}\in[-1,1]$ that represent the mean opinion of 
agent $j$ on a large set of issues or, considering that the information exchange is much faster than the adaptation dynamics,
 the opinion agent $j$ has about the mean issue. The mean issue is objective and it is described by a set of $D$  
numbers $\mathbb{Z}\in \mathbb{S}^D$. 

The relevant variables, representing the society, are the  internal variables of the agents.
Every agent $j$ has two main  properties. (A) Its internal state is determined by a set of  $D$ weights   
$\mathbb{J}_j$ ({\it moral vector}), which is invisible to other agents. (B) The main hypothesis in this work is that
while weights jointly code for prior experience,  they are subject to change due to the social interactions through 
a learning mechanism.

The vector $\mathbb{Z}$ changes in time reflecting social changes in moral parsing or values. We, however, consider that 
the adaptation dynamics of $\mathbb{J}_j$ is much faster than the dynamics of $\mathbb{Z}$ and suppose the 
latter as being fixed. We also concentrate on  $D=5$, but it  might be interesting  to explore the consequences of using different values.

The only interaction among agents comes from learning about the opinion fields 
of agents in their social neighborhood.  Learning occurs in 
order to decrease the psychological discomfort due to dissent. 
Learning is described by
a noisy gradient descent dynamics on a  potential function
 describing a psychological cost of disagreement
with each of its social neighbors.
In \cite{Cavi10}
we  have introduced a potential to model this and called it the psychological cost,
which depends on a parameter $\delta$, taking values between $0$ and $1$. 
$\delta$  represents an attempt to model
different cognitive strategies with respect with how novel 
or corroborating  information
is used in the learning process.
For $\delta=0$, as will be seen below, the agents
can be called error correctors. They only learn
from social neighbors from which they disagree by  bringing a  different
average opinion on the set of issues under discussion. Thus we also refer to these agents as {\it 
novelty} seekers, for they do not learn unless the information carries
a new and different point of view on the issues. For $\delta=1$, learning occurs by extracting
correlations. These agents learn from neighbors always, independently of
agreement or disagreement. 
This led us to call them {\it corroboration} seekers. 

\begin{figure}[!ht]
\centering \includegraphics[width=0.5\textwidth]{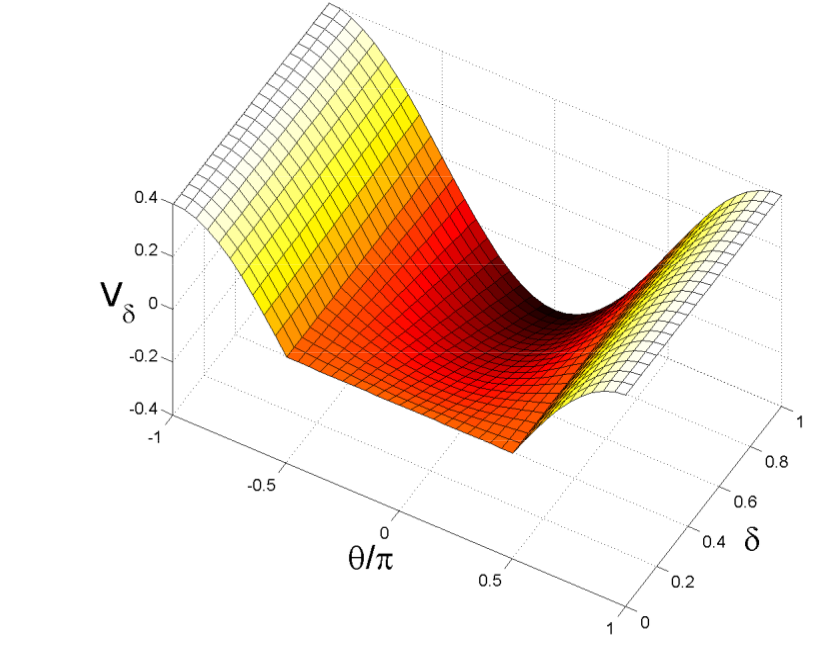}
\centering \caption{ {\bf Psychological cost function} $ V_\delta (h_i,h_j)$
for different values of $\delta$, as a function of 
$\theta_i/\pi$ , where 
$h_i
= \cos \theta_i$ 
for $h_j=0.4$ fixed. For fixed $\delta$, the
slope of the potential determines the scale of changes of the 
moral state vector. 
For $\delta=0$ changes only occur if there is a difference in 
the signs of opinion fields  $h_i$ and $h_j$. 
For $h_j>0$ this
occurs when $h_i$ becomes negative at $|\theta_i| > \pi/2$. 
On the
other extreme of cognitive styles, for $\delta=1$, any difference 
in magnitude of opinions has associated a slope of the cost function.
 }\label{fig:potencial}
\end{figure}

The family of psychological costs or interaction potentials, indexed by
$\delta$ (see figure \ref{fig:potencial}) is defined by:
\begin{equation}
V_\delta(h_i, h_j)= -\frac{1+\delta}{2} h_i h_j +
\frac{1-\delta}{2}|h_i h_j|
\label{socialcost}
\end{equation}
which can be written as 
 $V_\delta= -\delta h_i h_j$ for same sign opinions 
and  $V_\delta= - h_i h_j$ for opinions of a different sign.
We can also consider the total cost for a group homogeneous in $\delta$ leaving 
on a social graph ${\cal G}$ as
\begin{equation}
{\cal H}=\sum_{(i,j) \in {\cal G}} 
V_\delta(h_i, h_j).
\label{hamiltonian}
\end{equation}
There are a few reasons that justify using the same $\delta$ for both  agents in each interaction. First there is 
evidence \cite{Spears} that people tend to interact more with
those of similar cognitive styles. Second we have tried in simulations
with different $\delta$'s in the same population and in different
social networks and the qualitative results are similar, showing 
the robustness of this approximation. Finally a third reason 
is that it simplifies the analytical 
mean field calculations which we  discuss herein.

The specific form of the potential is inspired in 
learning algorithms for linear classifiers \cite{Engel,Vicente}. A Hebbian
algorithm can  be considered to lead to learning 
from the use of  correlations in the input and output 
units. Information from a pair (issue, opinion) 
will be embedded with a strength 
independent of whether a prediction was correct or not.
A Perceptron algorithm, on the other hand works by error
corrections. If the prediction, on an example was correct, 
it will not do any changes. Changes will be made only when 
the prediction was incorrect. In this sense we can say that
a Hebbian algorithm learns both from corroborating 
and from new information. A Perceptron algorithm will only 
learn from new information.

Learning proceeds in the following way. We consider a discrete time dynamics. 
At each time step an agent is chosen and its weights are updated, if
there is no noise in the communication, using  a gradient descent dynamics:
\begin{equation}
\mathbb{J}_i(t+1)= \frac{\mathbb{J}_i(t)-\epsilon \nabla_{\mathbb{J}_i(t)}{\cal H}} 
                                               {\| \mathbb{J}_i(t)- \epsilon \nabla_{\mathbb{J}_i(t)}{\cal H}    \| }, 
\label{noiseless}
\end{equation}
where $\epsilon$ defines the time scale.

We can also consider the case where noisy exchange of 
opinions  might drive the update uphill and to describe this scenario we 
introduce  an inverse temperature $\alpha$ and
a Monte Carlo Metropolis dynamics \cite{Newman99}. As usual, then choose a
 $D$ dimensional vector  $\mathbf{u} $ drawn  uniformly 
on a ball of radius $\epsilon$.   
A trial weight vector is defined by
\begin{equation}
\mathbb{T}=\frac{\mathbb{J}_i(t) +\mathbf{u}}{\| \mathbb{J}_i(t) +\mathbf{u} \|} 
\end{equation}
and accepted as the new weight vector, 
 $\mathbb{J}_i(t+1)=\mathbb{T}$ if 
the social cost decreases: $\Delta {\cal H}:= {\cal H}(\mathbb{T})- 
{\cal H}(\mathbb{J}_i(t)) \le 0$. If  $\Delta {\cal H} > 0 $
the change is accepted with probability $ \exp \left(-\alpha\Delta {\cal H} \right)$.
This leads, after a transient, to a distribution
of states given by the Boltzmann distribution
$P_B(\{\mathbb{J}_i\}) \propto \exp\left( -\alpha {\cal H}\right)$.

Alternatively we can proceed by making  explicit the
hypothesis
that the average social cost characterizes 
macroscopic states and suppose
that the expected value $\mathbb{E}[{\cal H}]$ has a certain value.  
The distribution of probabilities for 
the moral vectors of the agents has to be 
chosen  from among those that satisfy the information 
constraint and makes the least amount of 
additional hypotheses. This is the natural framework 
of maximum entropy, which leads again to the Boltzmann 
distribution. 
The  parameter $\alpha$, which characterizes the noise level in the 
first approach, appears now as a Lagrange multiplier. It
can be  seen to determine
the scale in which changes in social cost, 
brought about by differences in opinion, are important. 
This justifies being called the { \it scale of peer pressure}.
Thus the scale of peer pressure is analogous to the noise
amplitude of the exchange of information and to an inverse temperature
in statistical mechanics. If there is a high level of noise, then
the opinions of others will not be very influential, thus a low peer 
pressure. As the dependence of fluctuations on temperature permits
measurements of one in terms of the other, in statistical mechanics, 
fluctuations in opinions may be used to characterize the level 
of peer pressure in a society.

\begin{figure}[!ht]
\includegraphics[width=0.5\textwidth]{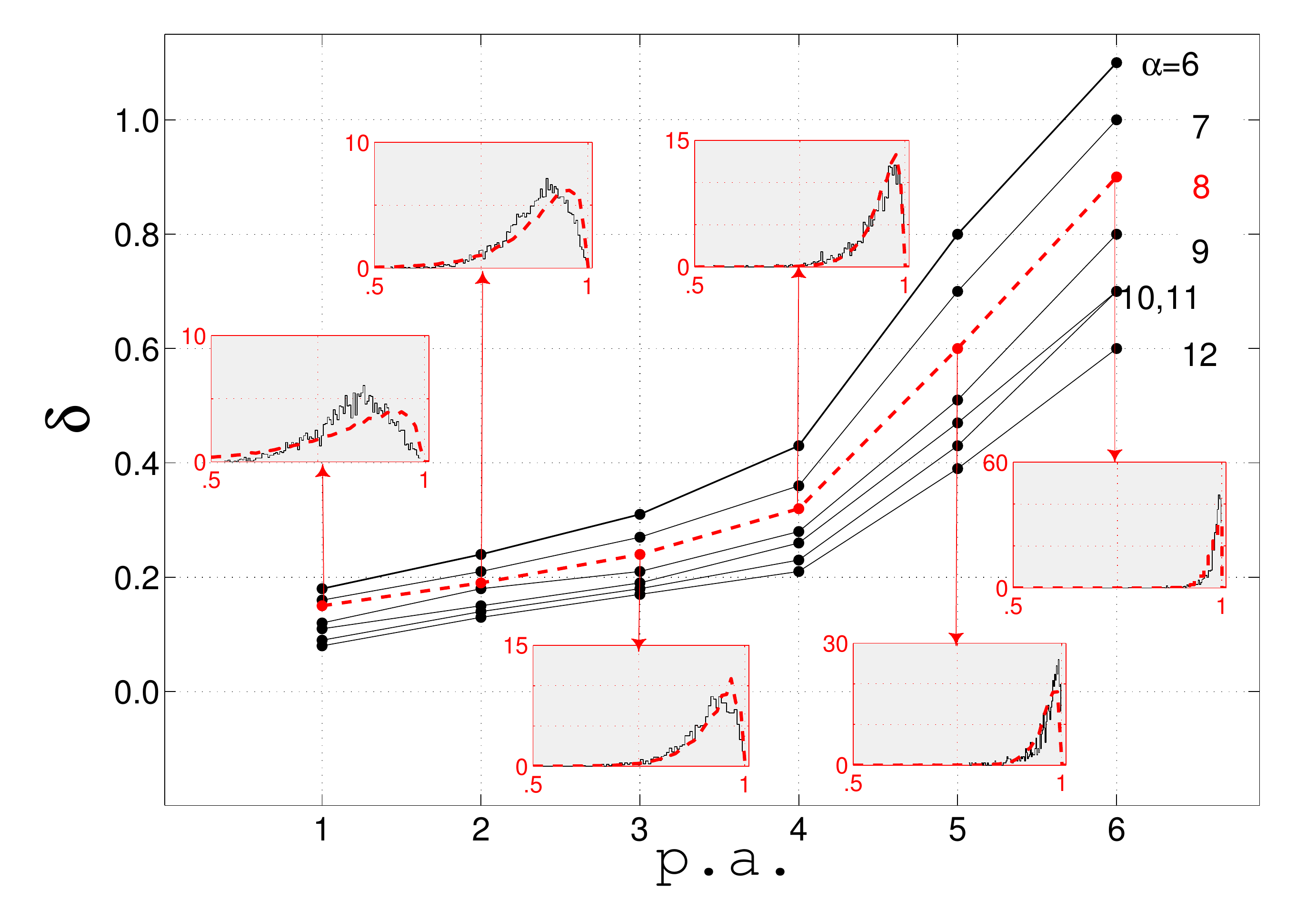}
\caption{{\bf Cognitive style and political affiliation}. The similarity of experimental and simulated  histograms 
suggests a link between cognitive style and political affiliation. In this figure $\delta$ is chosen at a fixed peer 
pressure $\alpha$ to yield a simulated histogram that matches empirical histograms. The insets show empirical histograms 
(black full line) and simulated histograms (dashed red line) for the case $\alpha=8$. The social graph is a Barab\'asi-Albert (BA) network \cite{barabasireview} with average degree $\bar{k}=22$ (branching parameter $M=11$) and size $N=400$. The horizontal  location of the insets
 is indicative of the political affiliation group to which the empirical histogram pertains. The best $\delta$ are depicted by the lines for several $\alpha$. The case $\alpha=8$ is indicated as a red dashed line.  The relation suggests that
 liberals have smaller $\delta$ than conservatives, meaning that liberals rely less on corroborating information that conservatives.  
 }\label{fig:delta_pa}
\end{figure}

This system has been studied  using Monte Carlo methods and 
Mean Field methods. The main empirical finding we focus on is the difference  in the number 
of moral foundations between self declared liberals and
conservatives. To study this we 
need to introduce an appropriate order parameter. 
 Our model has no semantics. Concepts like ``pure'', ``harmless'', ``loyal''
 in our model are just represented by indistinguishable dimensions of a vector space.
The  possibility of rotating  the frame of reference shows that any
initial interpretation  
of the coordinates is meaningless. But the set of issues under discussion 
introduces a symmetry breaking direction $\mathbb{Z}$, that may be regarded as 
the simplest vector to  characterize the society and what its
members are discussing.
We take this to be the direction parallel to the vector where all
coordinates are $1$, where all moral foundations are present and 
considered important. Our strategy then is to 
characterize both  the state of the agents model and 
 the empirical questionnaire
data  by introducing rotationally invariant order parameters 
which are semantically free as far as possible. 
Further analysis of the semantics of the model would 
be outside the present scope of this paper.

\begin{figure}[h]
\centering \includegraphics[width=0.5\textwidth]{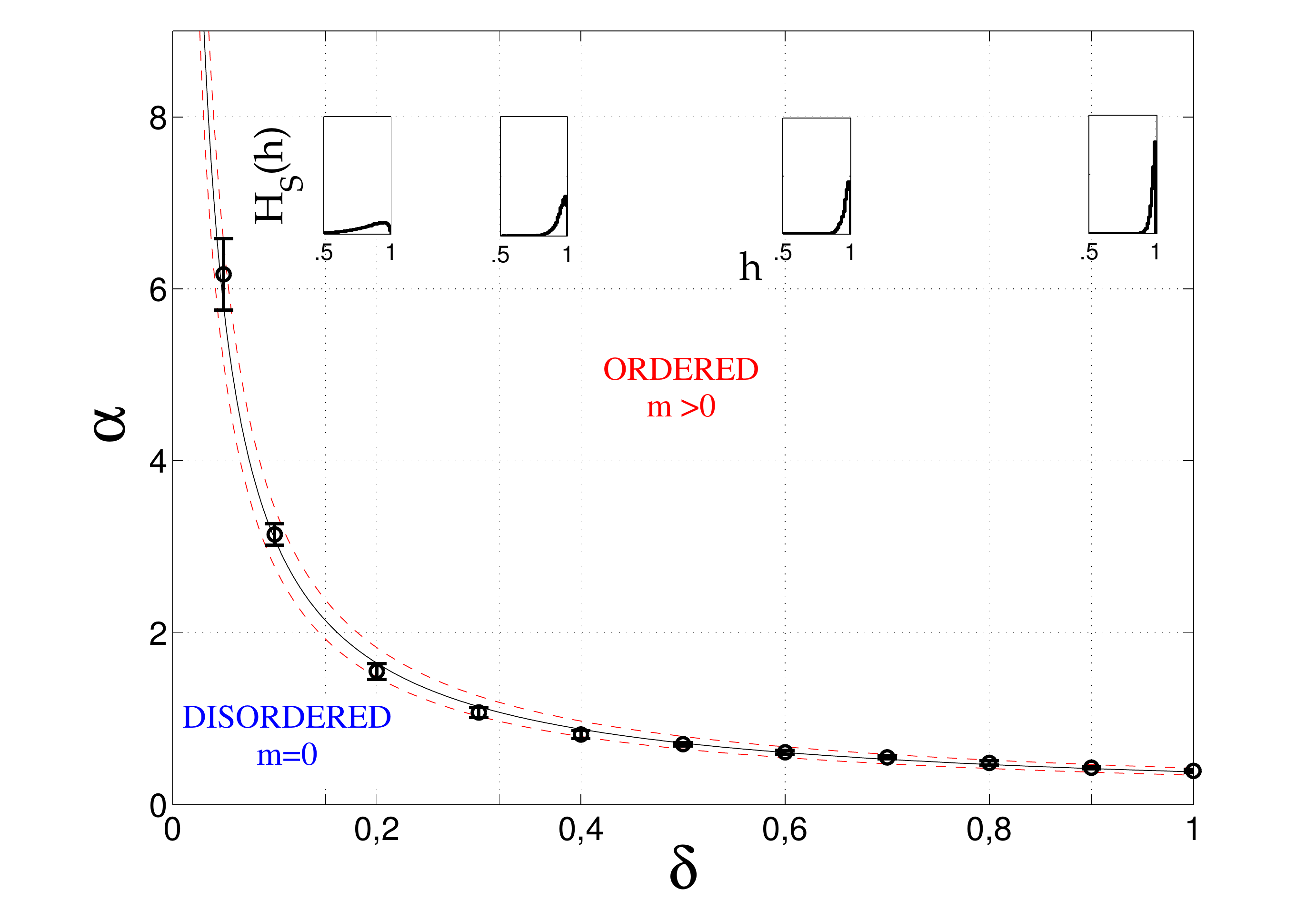}
\centering\caption{ {\bf Phase diagram in the space $\delta$ vs $\alpha$.} 
The case depicted corresponds to a BA network with average degree $\bar{k}=22$ and size $N=400$. Points correspond to $20$ runs of a Wang-Landau algorithm. The insets show the histogram $H_S(h)$ obtained as the equilibrium of a
Monte Carlo Metropolis dynamics with peer preassure $\alpha=8$ and $\delta$ provided by the optimization process that yields Figure \ref{fig:delta_pa}. The location of the insets is indicative 
of the parameters  $\delta$ and $\alpha$ used in the simulation. Full line indicates a fit $\alpha\propto1/\delta$, with red dashed lines corresponding to $95\%$ confidence error bars.
 }\label{fig:diagramadefase} 
\end{figure}

Simulated histograms $H_S$ for opinion fields $h$ are calculated as equilibrium distributions in a  Monte Carlo Metropolis dynamics run in a social graph that we here choose to be a Barab\'asi-Albert (BA) network \cite{barabasireview} with average degree $\bar{k}=22$ (branching parameter $M=11$) and size $N=400$. As the relation between the cognitive parameter $\delta$ and empirical affiliation $p.a.$ is unknown  we proceed by fixing the also unknown (but robust) peer pressure parameter $\alpha$ and finding, for each $p.a.=1$ to $6$ the $\delta$ that minimizes an Euclidean distance between simulated histograms $H_S(\ell \mid \delta,\alpha)$ and empirical histograms $H_E(\ell)$ defined  as
\begin{equation}
{\cal D}\left[H_S \mid H_E\right]= \sum_{\ell=-L}^{L}\left[H_S(\ell\mid \delta,\alpha)- H_{E}(\ell)
\right]^2,
\label{eq:dist}
\end{equation}
where the interval $[-1,+1]$ for $h$ is appropriately binned such that $h^{(\ell)}=\ell/L$ for $\ell=-L,...,L$. This procedure results in the curves depicted in Figure \ref{fig:delta_pa}. The model behavior is consistent with empirical evidence \cite{Amodio} in its general features, namely, $\delta$ is a nondecreasing function of the empirical political affiliation. We also notice the robustness in the qualitative behaviour as we vary the peer pressure $\alpha$. For the fits depicted as insets in the figure we use $\alpha=8$.

We also  calculate thermodynamic quantities by employing the Wang-Landau technique \cite{Cavi10,Wang01}. By doing that we are able to compute the phase diagram of Figure \ref{fig:diagramadefase}. From the point of view of ordering, the resulting diagram is straightforward exhibiting an ordered $m=\langle h \rangle >0$ phase and a disordered phase with  $m=\langle h\rangle =0$ separated by a continuous transition line that is well-fitted, in the case of a Barab\'asi-Albert network, by a simple power law $\alpha\propto 1/\delta$ \cite{Cavi10}.

\section{Mean field analysis}
\label{sec:mft}
This section aims at  providing theoretical support to the numerical results previously presented.
Our understanding of the model can be enriched by studying a tractable approximation 
with qualitatively similar behavior. Let us consider the set of issues to be  fixed 
(quenched disorder). In the analysis of this section we will not deal with the 
difficult task of averaging over quenched disorder, since it would 
draw attention and direct energy to technical issues beyond our current purpose. We
fix a set of issues and study the resulting
thermodynamics. The problem is still not simple and an exact solution
for the statistical mechanics problem is not known.  Here
we present mean field results obtained from information 
theory considerations in the form of a  Maximum Entropy argument.
We introduce a space of  tractable probability distributions,  
which factor over groups of
agents. The first and
 simplest choice is to consider a tractable family that factors
over the individual agents:
$
P_0 = \prod_i P_i
$. The parametrization of $P_i$ will be done in terms of the order
parameters which we still do not know. An advantage
of the mean field approach along these lines is that it tells
the relevant order parameters.
Our problem is reduced to minimization of the relative entropy
\begin{widetext}
\begin{equation}
S\left[P_0 \|P_B \right]=-\int  \left(\prod_i d\mu(\mathbb{J}_i)\right)
P_0
\ln \frac{P_0
}{P_B}
-\lambda (\langle P_0\rangle_\mu-1) - \alpha \langle\mathcal{H-E}\rangle_\mu 
\label{maxent}
\end{equation}
\end{widetext}
where ${P_B}$ is the Boltzmann distribution for the 
above Hamiltonian, $P_B= \exp (-\alpha {\cal H})/Z$ 
and $d\mu(\mathbb{J}_i)$
is the uniform measure over the surface of the $D$ sphere.
The relevant constraints that have to be taken into account
are normalization and that the expected value
 $\langle{\mathcal H}\rangle_\mu$ has a given fixed value $\mathcal{E}$, which might even be 
unknown, but is important in characterizing the state of
the agent society at least with respect to the opinions about the
issues.

\begin{widetext}
\begin{figure*}[!htb]
\centering
\includegraphics[width=0.7\textwidth]{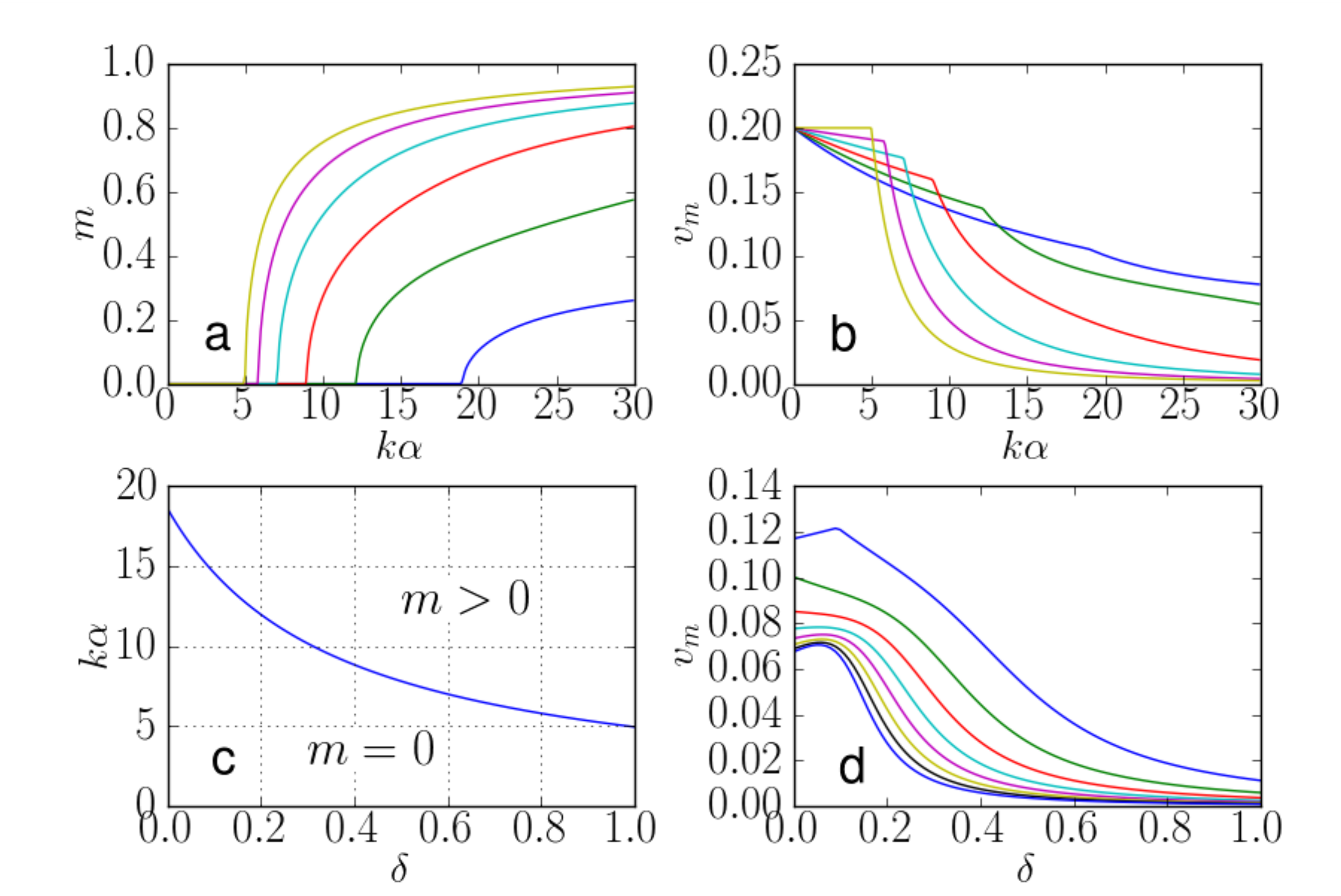}
\centering \caption{{\bf Mean field theory} : (a) $ m= \langle h \rangle$  as a function of total peer pressure $k\alpha$ for $\delta=0,0.2,0.4,0.6,0.8,1.0$ (from top to bottom) (b) $v_m=\langle h^2
  \rangle - \langle h \rangle^2 $ as a function of $k\alpha$ ($\delta$ from bottom up in the right side of the picture). (c) Phase diagram. (d)
width of distribution $v_m$ as a function of $\delta$, for fixed $k\alpha=10$ (circles) and $11$ (triangles).}
\label{fig:mag1ex}
\end{figure*}
\end{widetext}

We can drop the logarithm of the original partition function $\ln Z$ without 
changing  the
variational problem to obtain, from equations (\ref{socialcost}) and
(\ref{maxent}) 
\begin{eqnarray}
S\left[P_0 \|P_B \right]&=&-\sum_i \int  d\mu(\mathbb{J}_i)
P_i
\ln P_i - \lambda \int   d\mu(\mathbb{J}_i) P_i 
\nonumber \\
 &-&
\alpha
\sum_{(i,j)} \int   d\mu(\mathbb{J}_i)d\mu(\mathbb{J}_j)
 P_i P_j
 V_\delta(h_i,h_j)   \nonumber
\end{eqnarray}
and considering variations of the  set of $P_i$,
$
\frac{\delta S\left[P_0 \|P_B \right] }{\delta  P_i}=0
$,  leads to
\begin{eqnarray}
0&=&-1 -\lambda- \ln P_i -\alpha
\sum_{(j), \nu} \int   d\mu(\mathbb{J}_j) P_j
 V_\delta(h_i,h_j)   \nonumber 
\end{eqnarray}
This is an expression relating the probability density of an agent to those
of the social neighbors:
\begin{eqnarray}
P_i &\propto & \exp \left( -
\alpha
 \sum_{j} \int   d\mu(\mathbb{J}_j) 
P_j
V_\delta(h_i,h_j) 
\right) 
\end{eqnarray}
Now we go back to the problem of choosing the family of distributions
$P_i$. The main reason to call a family tractable is that the set
of equations above is closed.  
Depending on the structure of the Hamiltonian, different families can
be used.

The form of the Hamiltonian imposes the use of two order parameters
for each issue, which in order to close the set, we take to be
independent of the agent, 
\begin{equation}
\int   d\mu(\mathbb{J}_j)
P_j
V_\delta(h_i,h_j) =-\frac{1+\delta}{2} h_i m
 +\frac{1-\delta}{2} |h_i| r
\end{equation}
where we have introduced
\begin{eqnarray}
m& =& \int   d\mu(\mathbb{J}_j) P_j \;
\mathbb{J}_j \cdot \mathbf{x}  \label{eqm}
\\
r &=& \int   d\mu(\mathbb{J}_j) P_j \;
|\mathbb{J}_j\cdot \mathbf{x}|
 \label{eqr}
\end{eqnarray}
In principle the order parameters $m$ and $r$ could
have an index $j$ identifying the agent, but we make 
a reasonable assumption of homogeneity. This does not
mean that all agents are equal, but that they will
present values of the moral vector $\mathbb{J}_i$ 
drawn  from the same probability distribution.
Then the mean field 
probability distribution is given by
\begin{eqnarray}
P_{\mbox{\tiny MF}}(\{\mathbb{J}\}|k\alpha, \delta,m, r)
&=& \prod_{i}P_{\mbox{\tiny MF}}(\mathbb{J}_i|k\alpha, \delta,m, r) \\
&=& \prod_{i}\frac{\exp \left\{
 k \alpha
\left( \frac{1+\delta}{2} h_i m
 -\frac{1-\delta}{2}  | h_i| r\right)\right\}}{{\cal Z}_i},\nonumber
\label{PMFT}
\end{eqnarray}
where the denominators $\prod_i {\cal Z}_i$ ensure normalization
and  $k$ is the number of social neighbors.
Now equations \ref{eqm} and 
\ref{eqr} can  be seen not as the definitions of $m$ 
and $r$, but as the self consistent
mean field theory equations
from which their values can be calculated.

\begin{widetext} 
\begin{figure*}[!ht]
\centering \includegraphics[width=0.7\textwidth]{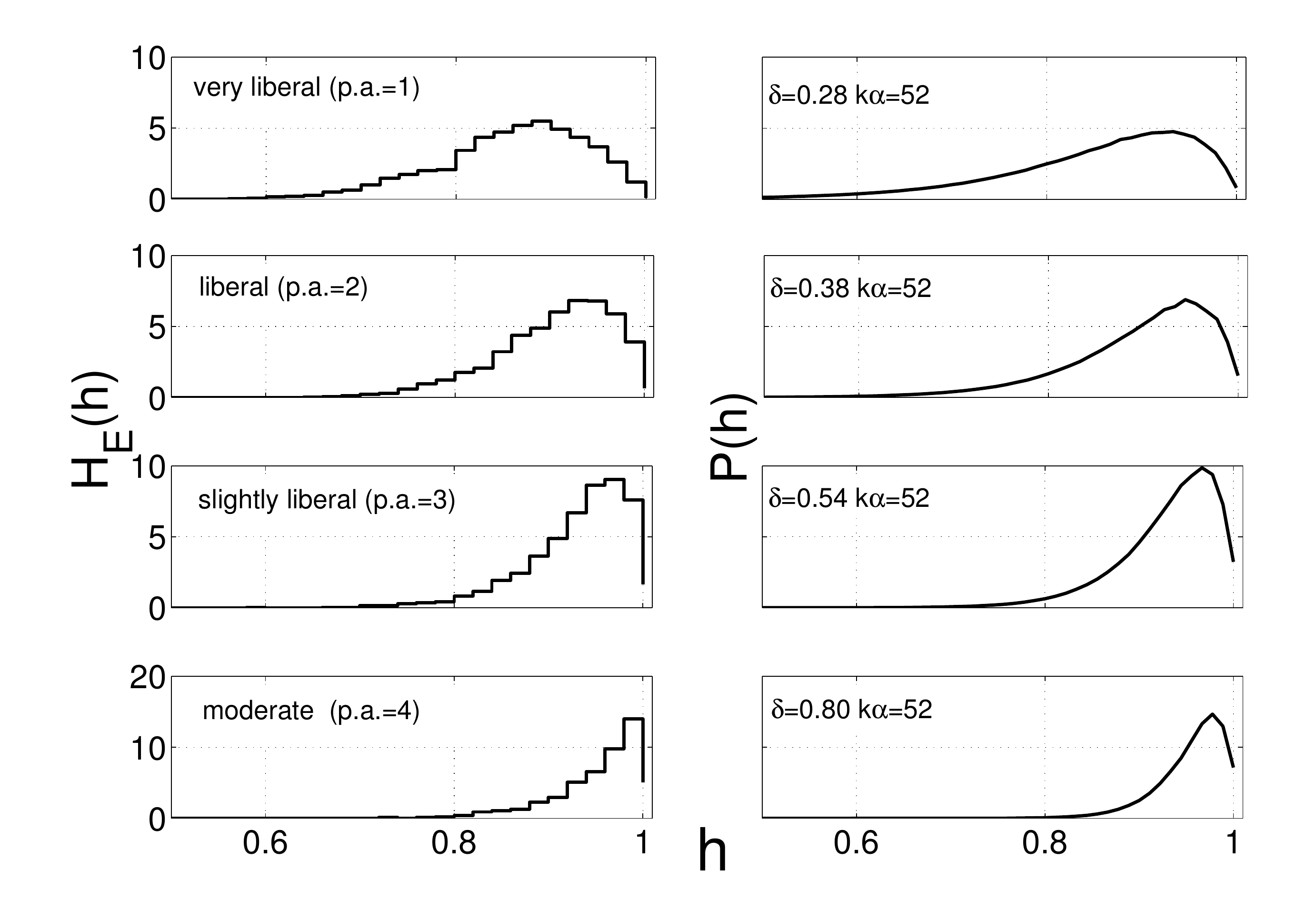}
\centering \caption{{\bf Mean field theory histograms}. 
Left column: Empirical histograms. Right column : mean field results. To recover histograms for $p.a.>4$ a larger $k\alpha$ is required. Vertical axes are shared in each row.
 }\label{fig:histogramMF}
\end{figure*}
\end{widetext}

The model can be studied for any value of the dimension of the
internal space, $D$. We use $D=5$ and our problem is reduced to doing some
integrals of up to five dimensions.
Since there is only one symmetry breaking direction $\mathbb{Z}$, 
we rotate the coordinate system such that 
$\mathbb{Z}$  is in the $\hat{e}_5$ direction. The polar angle
$\theta_3:=\theta$ with this direction is the only non trivial integration
variable since the 
other angular variables ($\theta_0, \theta_1 $ and $\theta_2$) drop
out and are trivial.

Call
$B(\theta|k\alpha,\delta,m,r):=\exp\left\{ k\alpha
\left( a m\cos \theta - b r |\cos \theta| \right) \right\}$
where $a:=\frac{1+\delta}{2}$  and $b:=\frac{1-\delta}{2}$, then 

\begin{eqnarray}
m&=& \frac{1}{{\cal Z}}\int_0^\pi
 d\theta \sin^3 \theta \cos \theta B(\theta|k\alpha,\delta,m,r)
\nonumber\\
r&=& \frac{1}{{\cal Z}}\int_0^\pi d\theta \sin^3 \theta |\cos
\theta| B(\theta|k\alpha,\delta,m,r)
\nonumber\\
{\cal Z}&=& \int_0^\pi
 d\theta \sin^3 \theta B(\theta|k\alpha,\delta,m,r)
 \label{zp1}
\end{eqnarray}
Equations  \ref{zp1} can be solved numerically self
consistently. Results in Figure  \ref{fig:mag1ex}a 
show the fixed points $m$ as a function of the total peer pressure, showing the existence of a 
phase transition as the critical line of total peer pressure 
 $k\alpha_c(\delta)$ depicted in \ref{fig:mag1ex}c   is crossed. 
The  critical  total peer pressure $k\alpha_c$ decreases with
larger values of $\delta$.  Figure \ref{fig:mag1ex}b shows the
width of the distribution of overlaps  (denoted $v_m$). An important
 prediction of the theory is that this depends strongly on the
 corroboration parameter $\delta$ as it can be seen in Figure \ref{fig:mag1ex}d.

We can use equation \ref{PMFT} to calculate the distribution 
of opinions about  the symmetry breaking direction  $\mathbb{Z}$
\begin{equation}
P(h|k\alpha, \delta)= \int d\mu(\mathbb{J}) 
\delta (\mathbb{J}\cdot\mathbb{Z} - h)
P_{\mbox{\tiny MF}}(\mathbb{J}|k\alpha, \delta, m, r)
\end{equation}
This is a mean field prediction that can be confronted to Monte Carlo
simulations and, more importantly, to experimental data. 
The result is
\begin{equation}
P(h|k\alpha, \delta)=\frac{1}{C}(1-h^2)\exp\left\{ k\alpha
\left( a h m - b r |h| \right) \right\}
\end{equation}
where ${C}=\int_{-1}^{1}(1-z^2)\exp\left\{ k\alpha
\left( a z m - b r |z| \right) \right\}dz$, is given to good approximation by
\begin{eqnarray}
C&=&\frac{2e^{\delta\tilde{m}}}{\delta^2 \tilde{m}^2}\left(1-\frac{1}{\delta\tilde{m}}\right)
 -\frac{1-\delta}{\delta \tilde{m}} 
+\frac{2(\tilde{m}-1)}{\tilde{m}^3}e
^{-\tilde{m}}
\end{eqnarray}
where $\tilde{m}= k\alpha m$ and we used that in the experimentally
relevant region $m=r$. Approximately
\begin{equation}
P(h|k\alpha, \delta)=\frac{(\delta\tilde{m})^2}{2}(1-h^2)e^{-\delta \tilde{m}
(1-h)}
\end{equation}

This comparison is shown in 
 Figure \ref{fig:histogramMF}, it hints  that the similarity
of the data and theory point to a relation between $\delta$
for the agents and political affiliation for the experimental subjects. 
Histograms for the more conservative groups resemble more the histograms
for agents with higher $\delta$'s; i.e. conservative behavior is more 
likely to be identified with larger reliance on corroboration and alternatively liberal behavior,  with smaller reliance. 

\section{Facebook network}
\label{sec:facebook}

In the previous sections we have based our discussion on simulations run on regular and ``synthetic'' BA networks. In this section we present simulation results on a realistic network extracted from Facebook network data \cite{traudetal,traud2} \footnote{Data can be downloaded at
\href{http://archive.org/details/oxford-2005-facebook-matrix}{http://archive.org/details/oxford-2005-facebook-matrix}}.

\begin{widetext}

\begin{figure*}[!ht]
  \centering \includegraphics[width=\textwidth]{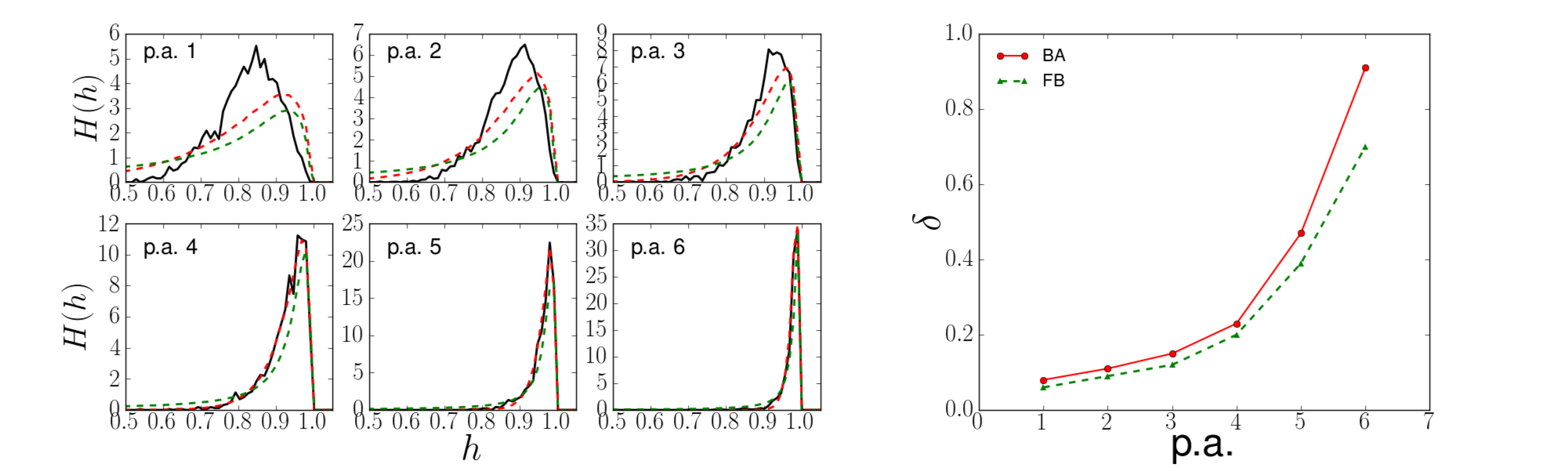}
  \centering \caption{{\bf Histograms for empirical data, synthetic
      and Facebook networks}. Left panel: The thick black lines are the empirical histograms $H_E(h)$, dashed red lines are the simulated histograms $H_S(h)$ with a Barab\'asi-Albert construction of size $N=800$ and $\bar{k}= 22$. The peer pressure is fixed to $\alpha=8$. The thin green line are the simulated histograms $H_S(h)$ using Princeton's Facebook network ($N=6596$ and  $\bar{k} = 89$) with $\alpha=1.98$. Right panel: Best $\delta$ as a function of $p.a.$ for the BA network (dashed red lines) and for Princeton's Facebook network (thin green line). }
\label{fig:paall}
\end{figure*}

\end{widetext}

In figure \ref{fig:paall} we show a comparison between empirical  $H_E(h)$ histograms and the best fit, in terms of  a the Euclidean metric, for the Princeton graph (size $N=6596$ and average degree $\bar{k} = 88.9$) and for a Barab\'asi-Albert construction with $N=800$ and $\bar{k}=22$ \cite{barabasireview}. To build theoretical histograms we have run Metropolis simulations \cite{Cavi10} fixing $\bar{k}\alpha=176$ in both scenarios and choosing, for each $p.a$, a $\delta$ that minimizes the metric defined by Eq. \ref{eq:dist}.  

As it is suggested by the mean-field approximation of the previous section, histograms only depend on the social graph topology through the average degree $\bar{k}$. Also we see that  to find the peer pressure per social neighbor $\alpha$  we have first to  measure the average degree independently.

\section{Dynamics: what do {\it conservative } agents conserve?}
\label{sec:dyn} 
\begin{widetext}
\begin{figure*}[!ht]
\centering \includegraphics[width=0.9\textwidth]{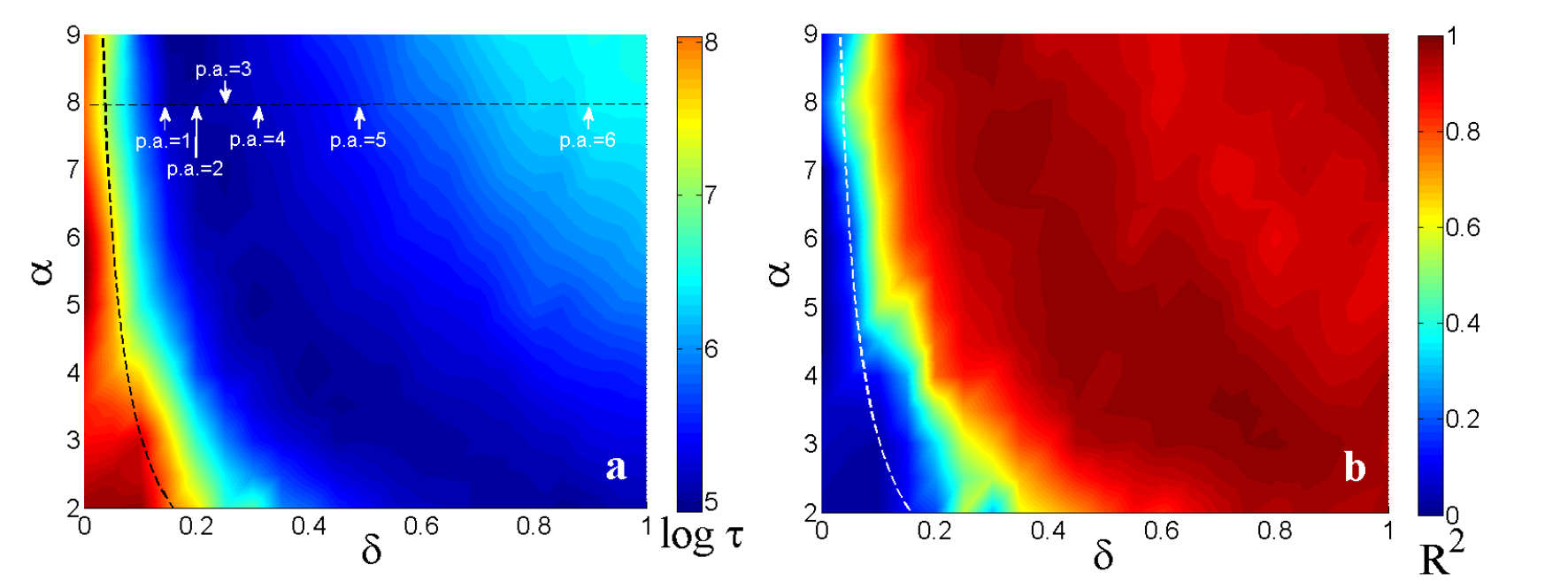}
\centering \caption{{\bf Characteristic re-adaptation times}. The topology is the same used to produce figure \ref{fig:diagramadefase}. Dashed lines
 indicate the phase transition. A rotation by an angle of  $\zeta=0.6 \pi$ rad has
 been employed as a perturbation to the Zeitgeist vector $\mathbb{Z}$ all over, however, conclusions are general.
(a) Re-adaptation times $\tau$ as a function of $\alpha$ and $\delta$. Arrows indicate political affiliation scores as
 identified in figure \ref{fig:delta_pa}, from $p.a.=1$ (very liberal) at left to $p.a.=6$ at right. Note that the minimum time occurs at 
values of $\delta$ inside the ordered phase (see figure \ref{fig:diagramadefase}) and that 
this value of $\delta$ corresponds to liberal and very liberal scores (figure \ref{fig:delta_pa}). (b) Re-adaptation times $\tau$ are inferred by 
regressing $\log {\cal D}$ against $t$. $R^2$ statistics are indicative that the data fits employed are only relevant inside the ordered phase.}
\label{taudelta}
\end{figure*}
\end{widetext}

In addition to obtaining that novelty
seeker agents are identified with  liberals and that corroboration
seeker agents are more similar to conservatives, the model 
can be studied to determine dynamic collective properties. In particular
we study in this section how groups of 
agents identified with  conservative or liberal
differ in time scales to adopt new positions. Given the relation between political affiliations and  the corroboration parameter suggested by the model, it would  be a contradiction if characteristic reaction times
to changes in the symmetry breaking direction $\mathbb{Z}$ turn out to decrease with
increasing  $\delta$. 
So, putting the theory to the test,   we now turn to study 
the response to changes of the issues and how 
the group accommodates to such changes.
Once the MC simulation has equilibrated, we change the $\mathbb{Z}$. The
new direction and the old one have an overlap  $\mathbb{Z}_{old} 
\cdot \mathbb{Z}_{new}=\cos\zeta$. We continue the  Metropolis simulation
and characterize, as a function of simulation 
time, the distance to the equilibrium distribution. 
A natural distance  from equilibrium would be a measure
of the Kullback-Leibler divergence. However, we do not have access
to a theoretical form of the out-of-equilibrium distribution.
A simpler procedure is to calculate a distance directly
from the histograms. After a MC step, which includes a learning sweep
over the whole population, we obtain $H_{t}(h)$
the histogram of opinions 
$h_{new}=\mathbb{J}\cdot\mathbb{Z}_{new} $
about the new symmetry breaking direction, giving the fraction
of agents with opinion in a given range. Define the Euclidean distance by
\begin{equation}
{\cal D}\left[H_{t} | H_{eq}\right]= \sum_{h=-1}^{1}\left(H_{t}(h)- H_{eq}
(h)
\right)^2
\label{eq:dist2}
\end{equation}
where the range of the variable $m_Z$ has been discretized into $20$ bins.
The distance from equilibrium as a function of time can be
parametrized as ${\cal D}(t)= F(\zeta) e^{-t/\tau}$, 
where the measured $\tau=\tau(\zeta,\alpha,\delta)$ appears in figure \ref{taudelta}.
The valley, shown
in blue, shows the region where the agents are faster
to re-equilibrate adapting to the new conditions. It  occurs inside the ordered 
phase, not in the high $\delta$  region of the conservatives, nor at 
the border of the phase transition. This is to be expected, since at the border
there is critical slowing down. The interesting thing is that 
the group of agents that 
re-adapts to equilibrium the fastest is the one which
has been identified with the most liberal subjects of the data.

The surprise lies not in that ultra-liberals adapt the fastest, but
that our simple model is also consistent in this respect. 
This result is central to our proposal for interpreting  what conservative and liberal
means in a group of agents. Agents with high $\delta$ are conservative
and with lower $\delta$ more liberal, also in terms of their time to
adapt to changes. Based on this we can attribute political labels
to the agents which are consistent
with the attribution based on the data since
they show the same dependence on $\delta$.

\section{ Discussion: Diversity of cognitive styles }
\label{sec:discussion}
Statistics and reaction times permit identifying different
cognitive styles with different effective dimensions or opinions
about a mean issue (that we call {\it Zeitgeist} in \cite{Cavi10}). 
Therefore given that people do present 
different cognitive styles and that they interact and learn from each other
we expect  
that there will be people who hold different  sets of moral values.
Cultural wars will follow from diversity of cognitive styles.

This is a semantic free conclusion. The model cannot distinguish
between the different moral foundations. There must be another 
reason why some of the foundations are always present while others 
may be absent. Evolutionary arguments by Haidt 
go a long way in explaining why the harm/care and justice dimensions 
are more uniformly common. They may be found, to some extent in other
primates \cite{DeWaal,Katz}. The emergence of the other dimensions, which seem 
to be present only in humans, are supposed to foster higher
cooperation levels and ultra-social behavior. Suppose, as we do, 
these facts to be
reasonable, then the question that emerges is why society has kept
all types of cognitive styles and not only those that lead to a more
cohesive society? A possibility is that different strategies within 
a society are useful to  cope collectively 
with different challenges. Higher
cooperation level gives higher fitness during times where current opinions
lead to correct answers from a survival point of view. Conservative
behavior would then be the fittest when maintaining current behavior is beneficial to 
the society. However, during times when current opinions are not guiding
in the finding of useful answers, in a  survival enhancing sense,
a different perspective is needed. A larger menu of choices
may permit finding better alternatives in an efficiently distributed
manner. 

When the Zeitgeist changes, due to external conditions, or
due to a new issue being introduced to the debate, a more liberal 
approach seems reasonable. The question, from the current perspective,
is then translated to whether this behavior can be seen within our model.
In figure \ref{fig:delta_pa} we presented a connection between agents 
characterized with a given $\delta$ parameter and the political 
affiliation of the questionnaire respondents. A question that
arises is why a lower but nonzero 
value of $\delta$ was found
for a peer pressure around $\alpha =10$ which is
well inside the ordered phase? Why were not the ultra-liberals associated
to a $\delta$ right on the edge of the phase transition? Maybe the reason 
can be found in the dynamics of adaptation to the new Zeitgeist. We find
from the simulations, within the appropriate $\alpha $ region,
 that at that value, $\delta \approx 0.20 - 0.35$  the characteristic time of
re-adaptation to a new Zeitgeist has a minimum
(see figure \ref{taudelta}). The lower the conservatism of a population 
the less cohesiveness it will present 
in responding to external challenges as a group. There is no benefit
in being more liberal than necessary. Ultra-liberals are not
on the disordered phase, but in the ordered phase. They even 
 are not at the border of the transition, they are in a way prevented
from being on the  disordered phase by critical 
slowing down. Closer to the border the
system is softer but takes longer to rearrange. And from our results it
seems that even the ultra-liberals observed in the data rely on some corroboration in order
to construct their moral vector.

\section{\label{sec:conclusions} Conclusions and perspectives}
The modeling of social systems has a long (and well-fought \cite{GalamBook}) history. It might 
be surprising to some that a mathematical model can be constructed
and  directly confronted to data, replicating some statistical findings and 
making predictions borne out by observations. We believe that this is possible
by setting the problem in the context of information theory. 
After relevant variables were identified, information about
neurocognitive, psychological and social science was used 
to attribute a probability distribution for the variables, 
which is finally used to estimate relevant
experimental signatures from order parameters. 
This is, ultimately, what is done in traditional areas of Physics.

We have presented results from Monte Carlo numerical methods 
and analytical approximation schemes such 
as mean field for a model of interacting agents. 
These techniques are suited to study the collective or aggregate
properties of our model of agents.  Drastic changes in collective properties
signal phase transitions and the emergence of different
regimes of behavior. 

The neural networks of the agents are  quite simple. 
The only way to know if exaggerated simplifications
have been made is to compare with data. Even if not useful for
heuristic confrontation, models may be be useful in their own right as
laboratories where we develop intuition about the different
methodologies needed to extract information from possible more complex
models of the future. They help in formulating a set of questions that can be
addressed experimentally and theoretically. By pointing out their own
limitations,  current models can bring us closer to more useful models in
the future. The networks are not supposed to model the 
brain networks of individuals, but rather the fact that 
people integrate the different moral dimensions of an issue, weighted
by their own views about the importance of each dimension, in order
to reach conclusions in an intuitionist way rather than by using
a rationalist {\it  if-then}  set of rules. 
 
A summary of conclusions about our results should first of
all mention what we have not attempted to do. 
 No mention of any evolutionary
perspective of how the  moral foundations came to be
was presented. In particular it seems reasonable to agree about
the possible enhanced fitness that may derive from increased social 
capital of a more ordered society but
this   should be central in future discussions. If this is granted, then
we have to answer why lower social capital promoting cognitive
strategies should be present and not have been eradicated by selection.
Are liberals just free riders invading a society of authority/loyalty/purity
respecting conservatives? Trying to stay aside of semantic interpretations, 
we give an evolutionary reason why cognitive styles compatible 
with liberal behavior are found in modern times and 
have not been purged. 
Reaction times of the society of agents show that it is consistent
to call large $\delta $ agents as conservatives, since they
have a large equilibration time under changes of the Zeitgeist. 
On the flip side, small $\delta$ is expected to correlate with 
liberal fast adapting behavior under the same changes. But
this was common knowledge. 
What is the novelty of conservatives taking more
time to re-adapt than liberals? We found that liberals do not correspond
to a $\delta=0$ cognitive style. Not only conservatives, but also
liberals are on the ordered side of the phase diagram. But
as we approach the disordered phase, critical slowing down sets in. So
agents with $\delta$ too small will also have large equilibration
times and these have probably been eliminated.
 A compromise between being fast to re-adapt and having
high social capital shapes the societies of agents that live in an
ever changing environment.

We have shown that different cognitive styles give rise through 
social interactions to different statistics about the opinion field $h$.  
The interactions are represented by a potential
that although it was never intended to claim precise realism 
it captures several stylized features of human cognitive styles. 
We have been cautious to allow agents to learn from opposing views. 
While this may not always occur in human interactions, there 
are certain windows of time where people acquire their moral values
from their social network of interactions. 
Qualitative information from fMRI and psychological 
tests about cognitive properties have been used to construct 
the interactions but future work will have to refine
the learning algorithms. Agents in social networks have shown 
a better agreement with the experimental data than simulations
in e.g. square lattices and the model  successfully  predicts what sort of connectivity
 is to be expected if the subjacent social network is  complex.   We feel that it is rather remarkable that from 
the answers of questionnaires and by postulating certain information
exchange mechanisms something about the topology of 
the social interactions of a society can be inferred. 

Many questions are raised. While we are aware of the previous use of 
the term peer pressure, we have introduced a quantitative definition
that might lead to experimental characterization and measurement. 
This might help deciding whether our concept is useless or not, 
but it is the nature of experimental work to help decide 
relevance. An interesting consequence 
of our approach and of the idea of peer pressure is that
histograms of effective dimensions might change after 
external threats to a society are detected. From the 
properties of the model we can  predict that the  mean of histograms
$H(h)$ will increase
and variances of the distribution will be reduced after 
external threats are detected. 
 The model also predicts that societies that discuss a wider set of 
issues will move in an opposite direction, with a reduction of the mean of
$H(h)$ and an increase
of its variance.

 Several methodological problems are raised and
should be analyzed, among them,  the measurement
of the peer pressure, the parsing of moral discourse into $5$ or more 
dimensional vectors,  the determination of the Zeitgeist vector, time scales
of change. Among the theoretical topics, we should approach the
problem of semantics and dress the different moral dimension
 of the model with  an interpretation in the language of moral philosophers.
Evolutionary considerations will probably guide the process and break
the remaining symmetries. 
 We have also neither addressed a possible role of genetic factors influencing
cognitive styles nor if the value of $\delta$ depends on the agent's 
environment. For the latter we will have to consider more 
sophisticated learning algorithms. 
Understanding evolutionary and cognitive influences behind cultural wars
and their mathematical modeling seems to be a reachable goal.

\begin{acknowledgments}
We thank Jonathan Haidt for kindly permitting us to have access to his data on moral foundations. We also thank Eytan Domany for the permission to use the SPIN software. This work received financial support from FAPESP (grant 2007/06122-0), CNPq (grant 550981/07-1) and The Center for Natural and Artificial Information Processing Systems at The University of S\~ao Paulo (N\'ucleo de Apoio \`a Pesquisa CNAIPS-USP).
\end{acknowledgments}

\section*{References}
\bibliographystyle{apsrev4-1}
\bibliography{moral}

\end{document}